\documentclass[amsmath,amssymb,pra,aps,mathbbm,superscriptaddress]{revtex4}
\usepackage{amsmath,amsfonts,amssymb,amsthm,graphics,graphicx,epsfig,bbm}
\usepackage[colorlinks=true,citecolor=blue,linkcolor=blue,urlcolor=blue]{hyperref}
\usepackage[usenames]{color}
\usepackage{graphicx}
\usepackage{subfigure}
\usepackage{epsfig}
\usepackage{bbold}
\usepackage{dcolumn}
\usepackage{bbm}
\usepackage{color}
\usepackage{verbatim}
\usepackage{epstopdf}
\usepackage{amssymb} 
\usepackage{amstext}
\usepackage{latexsym}
\usepackage{hyperref}
\usepackage{amsfonts}
\usepackage{psfrag}
\usepackage{xcolor}
\usepackage{times}
\usepackage{mathtools}
\newcommand{\ketbra}[2]{\vert #1 \rangle \langle #2 \vert}

\newcommand{\ket}[1]{\ensuremath{\left|{#1}\right\rangle}}
\newcommand{\bra}[1]{\ensuremath{\left\langle{#1}\right|}}

\usepackage{mathptmx} 

\usepackage{subeqnarray}

\DeclareMathAlphabet\mathcal{OMS}{cmsy}{m}{n}

\begin{document}

\title{Degree of Quantumness in Quantum Synchronization}
\author{H. Eneriz}
\affiliation{Department of Physical Chemistry, University of the Basque Country UPV/EHU, Apartado 644, E-48080 Bilbao, Spain}
\affiliation{LP2N, Laboratoire Photonique, Num\'erique et Nanosciences,
Universit\'e Bordeaux-IOGS-CNRS:UMR 5298, F-33400 Talence, France}
\author{D. Z. Rossatto}
\email{dz.rossatto@unesp.br}
\affiliation{Departamento de F\'{\i}sica, Universidade Federal de S\~{a}o Carlos, 13565-905 S\~{a}o Carlos, SP, Brazil}
\affiliation{Universidade Estadual Paulista (Unesp), Campus Experimental de Itapeva, 18409-010 Itapeva, S\~{a}o Paulo, Brazil}
\author{F. A. C{\'a}rdenas-L{\'o}pez}
\affiliation{International Center of Quantum Artificial Intelligence for Science and Technology~(QuArtist)\\
and Physics Department, Shanghai University, 200444 Shanghai, China}
\author{ E. Solano}
\affiliation{Department of Physical Chemistry, University of the Basque Country UPV/EHU, Apartado 644, E-48080 Bilbao, Spain}
\affiliation{IKERBASQUE, Basque Foundation for Science, Mar\'{i}a D\'{i}az de Haro 3, E-48013 Bilbao, Spain}
\affiliation{International Center of Quantum Artificial Intelligence for Science and Technology~(QuArtist)\\
and Physics Department, Shanghai University, 200444 Shanghai, China}
\author{M. Sanz}
\email{mikel.sanz@ehu.eus}
\affiliation{Department of Physical Chemistry, University of the Basque Country UPV/EHU, Apartado 644, E-48080 Bilbao, Spain}

\begin{abstract}
We introduce the concept of degree of quantumness in quantum synchronization, a measure of the quantum nature of synchronization in quantum systems. Following techniques from quantum information, we propose the number of non-commuting observables that synchronize as a measure of quantumness. This figure of merit is compatible with already existing synchronization measurements, and it captures different physical properties. We illustrate it in a quantum system consisting of two weakly interacting cavity-qubit systems, which are coupled via the exchange of bosonic excitations between the cavities. Moreover, we study the synchronization of the expectation values of the Pauli operators and we propose a feasible superconducting circuit setup. Finally, we discuss the degree of quantumness in the synchronization between two quantum van der Pol oscillators.
\end{abstract}

\maketitle
\section{Introduction}
Synchronization is originally defined as a process in which two or more self-sustained oscillators evolve to swing in unison. The original intrinsic frequencies are modified by the interaction between the oscillators, and a common effective frequency is observed~\cite{Pikovsky2003, luo}. It is a rich phenomenon manifested in a variety of disciplines that was typically studied in classical settings. Examples beyond pendula include heart beats~\cite{Cai1993}, neural networks~\cite{Rosin2013} and beating of flagella~\cite{Goldstein2009}. 

During the last decades, a significant progress has been achieved in quantum technologies, which has allowed the search of synchronizing behaviors  in quantum platforms~\cite{Goychuk2006, Zhirov2006}. Since then, the most studied case consists of chains of quantum harmonic oscillators with driving fields, dissipative mechanics, and nonlinearities~\cite{Lee2013, Manzano2013, Walter2014, Lee2014}. These models can be straightforwardly compared with the classical case, which corresponds to the presence of many quanta in the model. Quantum mechanics, on the other hand, introduces two main effects, namely, quantum noise and quantum correlations~\cite{Giorgi2012}. Quantum correlations have been reported to be strong in quantum synchronization~\cite{Hush2015} and, indeed, a synchronization between micromasers stronger than expected by semiclassical models has been recently discussed~\cite{Davis2016}. Likewise, it is also reported that synchronization is closely related with the performance on quantum heat engine \cite{arXiv.1812.10082}. In such case, the relation between the bath temperatures imposes a bound on the performance of such system. Furthermore, measures to quantify the synchronization of continuous variable quantum systems, such as two coupled optomechanical systems and topological lattice of these systems, have also been proposed~\cite{Mari2013,arXiv.1908.07296}. Moreover, it has also demonstrated that synchronization it is closely related to a symmetry breaking on the system dynamics rather than the values of the macroscopic parameters of the model associated \cite{arXiv.1907.03323,arXiv.1907.12837} . Finally, in Refs \cite{arXiv.1909.12050,arXiv.1909.12703}  it has been reported the use of the synchronization on two-level system as a feature to reduce the computational complexity of the quantum key distribution protocol.

The study of quantum synchronization has also been extended to quantum systems without classical analogue, e.g. two-level systems~\cite{Zhirov2008, Zhirov2009, Giorgi2013}. The lack of a classical counterpart makes the definition of synchronization non-trivial, and it has been addressed by studying periodically oscillating observables, and recently has been addressed considering local dissipation \cite{arXiv.1907.06886,Phys.Rev.Lett.123.023604}. These approach have been further validated by measures of quantum correlations, such as quantum mutual information~\cite{Ameri2015}, and the first practical applications in qubits have been recently presented \cite{giorgi2016,bellomo2017,cardenas2017}. However, some of the aforementioned results might be considered as classical synchronization processes in quantum setups, as we will explain  below. Moreover, we will show that synchronization can occur even when there are no quantum correlations between the synchronized parts in the steady state~\cite{Ameri2015}, which has been also recently noted in Ref.~\cite{zam2016}. This rises the question about the quantumness of quantum synchronization processes.

In this Article, we address the problem of how quantum a quantum synchronization process is from the point of view of quantum information. To this aim, we introduce the concept of \textit{degree of quantumness} $\Xi$ of quantum synchronization. Afterwards, we illustrate it in two composed cavity-qubit systems, showing that one can tune internal parameters to achieve all possible degrees of quantumness between the qubits. Then, to exemplify it, we propose a feasible circuit quantum electrodynamics (cQED) setup. Finally, we briefly discuss the extension of the concept of degree of quantumness to infinite-dimensional quantum systems, as is the case of quantum van der Pol oscillators. It is worth stressing that we are not interested in developing another method to quantify how much two observables of two different systems are synchronized. Our goal here is the identification and quantification of the true quantum nature of a synchronization process between two quantum systems. For the sake of clarity, hereafter, when we assert that two observables are synchronized, we are considering the general case in which their dynamics converge to periodic oscillations with the same frequency. In this manner, these processes can be classified in any type of synchronization, e.g, in-phase, anti-phase or complete synchronization, and so on~\cite{Pikovsky2003,luo}. To characterize and to quantify the type of synchronization between two observables, one can make use of the already known measures of synchronization \cite{Pikovsky2003,luo,zam2016}.  

A natural language to deal with synchronization is information theory, since the parties share out information during this process. In this sense, the mutual information was proposed as an order parameter for signaling the presence or absence of quantum synchronization \cite{Ameri2015}. However, this quantifier is not sufficient to answer the question of {\it how quantum} this process actually is and, indeed, one could straightforwardly engineer quantum dynamics in which only one observable is synchronized. However, from the point of view of quantum information, this synchronization may be considered classical, since there exists an equivalent classical dynamics describing the same synchronization process. This approach has already been followed in the context of partial cloning of quantum information~\cite{Ferraro2005, Lindblad1999, Sanz2010} or bio-inspired quantum processes~\cite{Alvarez-Rodriguez2014}, and could be useful to quantify the quantumness of quantum operations with respect to the environment~\cite{Meznaric2013}. Along these lines, we extend this idea to quantum synchronization, constructing a quantifier of the quantumness of the process.

\section{Degree of quantumness}
We will formally define the concept of \textit{degree of quantumness} $\Xi$ of quantum synchronization for a bipartite system, since the extension to multipartite cases is straightforward. Let us consider a bipartite quantum system $\mathbb C^{d}\otimes \mathbb C^{d}$, which can also be coupled to a complex environment. Let $\mathcal {S} = \{A_{i} \in \mathcal {M}_{d}(\mathbb C) | A_{i} = A_{i}^{\dagger} \} {}_{i=1}^{\chi}$ be a set of all linearly independent operators which simultaneously synchronize in both subsystems, with  $\chi = |\mathcal {S}|$, that we call \textit{cardinality of quantum synchronization}, and the sets $X_{k} = \{ A_{i} \in \mathcal {S} | [A_{k}, A_{i}] = 0 \}$ and $c = \max_{k} |X_{k}|$. Then, the \textit{degree of quantumness} of quantum synchronization is given by $\Xi = \chi - c$. Notice that $0 \le \Xi \le d^2-d$ and that $\mathcal {S}$ is a set but not a vector space since, if $A, B \in \mathcal {S}$, it does not necessarily mean that $A + B \in \mathcal {S}$. The reason is that even though $A$ and $B$ synchronize, they could do it with different frequencies and phases, so that linear combinations do not synchronize in principle. Therefore, linear independence removes the redundancies when more than one operator synchronize with the same frequency. 

Let us remark that, if the degree of quantumness of a given synchronization process is $\Xi=0$, i.e. every operator synchronized in both subsystems can be diagonalized in the same basis, then this is just classical synchronization from the point of view of information theory. The reason is that an equivalent classical dynamics synchronizing for the same operators can be constructed \cite{Sanz2010}, since only populations (diagonal terms of the system density matrix) synchronize, with the coherences (off-diagonal terms that take into account the quantum superposition) remaining desynchronized. On the other hand, if the degree of quantumness is maximum, i.e. $\Xi=d^{2}-d$ , then all non-commuting observables are synchronized, that is, not only the populations but also the coherences are synchronized. In the case of synchronization with the same frequency, phase and amplitude is equivalent to the synchronization of the reduced density matrices. From the point of view of quantum information, we call this a complete quantum synchronization.

\begin{figure}[t!]
\centering
\includegraphics[width=0.5\textwidth]{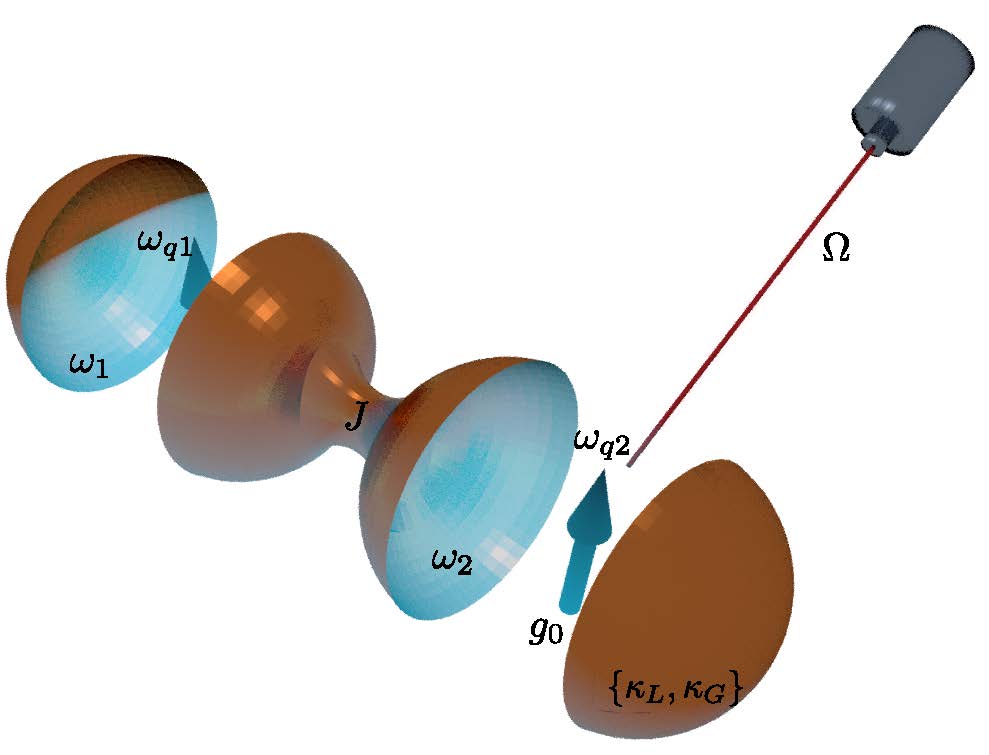}{\vspace{1em}}
 \caption{Quantum optical implementation of Eq.~\eqref{Eqn1}. The qubits, represented by arrows, are strongly coupled to the cavities, which can interchange photons coherently with rate $J$.}
 \label{Fig1}
\end{figure}  
Continuous-variable systems, i.e. infinite dimensional quantum systems, deserves an special mention. In this case, it is obvious that the number of linearly-independent Hermitian operators which are necessary to retrieve the information about the density matrix is infinite. For instance, a harmonic oscillator can be described through all its moments $\frac{1}{2}\langle x^k p^n+ p^n x^k \rangle$ and $\frac{i}{2} \langle x^k p^n- p^n x^k\rangle$, which are linearly independent \cite{barnett}. Indeed, in case of a total quantum synchronization, the cardinality is $\chi = |\mathcal{S}| = \infty$, which opens several theoretical challenges. A prototypical model is a network of van der Pol (vdP) oscillators. The quantum version of this model has recently attracted much attention, and several proposals for engineering it in the oscillating dynamics of trapped ions or nanomechanical oscillators has been put forward~\cite{Lee2013, Walter2014, Lee2014}. We will study below an example of degree of quantumness with quantum vdP oscillators.

\section{The model}
We illustrate now our definition in a setup consisting of two coupled cavities with a qubit in each of them, and investigate the synchronization between the two-level systems, in which the natural observables are Pauli operators. In our work, we are interested not only in one of the components of the spin operator~\cite{Giorgi2013,Ameri2015}, but in all of them. In this case, the maximum degree of quantumness happens when the expectation values of the three Pauli operators, $\sigma_{x}^{1}$, $\sigma_{y}^{1}$ and $\sigma_{z}^{1}$, are synchronized with their counterparts, $\sigma_{x}^{2}$, $\sigma_{y}^{2}$ and $\sigma_{z}^{2}$, respectively. The two cavity-qubit systems, depicted in Fig.~\ref{Fig1}, are coupled through a hopping term in the degrees of freedom of the cavities. A weak driving field acts on one of the two-level systems. Furthermore, the cavities loss and gain energy at rates  $\kappa_{L}$, and $\kappa_{G}$, respectively. The dynamics of such systems can be described by the master equation

\begin{eqnarray}
 \label{Eqn1}
\dot{\rho}=-i[\mathcal{H},\rho ] + \sum_{j=1,2} \big[\kappa_{L,j}\mathcal{D}[a_{j}]\rho + \kappa_{G,j}\mathcal{D}[a_{j}^{\dag}]\rho\big],
\end{eqnarray}
here, $\kappa_{L,j}$, and $\kappa_{G,j}$ correspond to the loss and gain energy rate, respectively. Furthermore, $\mathcal{D}[O]=O\rho O^{\dag}-\frac{1}{2}\{O^{\dag}O,\rho \}$ is the Lindblad superoperator. Finally, $\mathcal{H}$ is the system Hamiltonian given by the following expression
\begin{eqnarray} 
\label{Eqn2}
\mathcal{H} =  \sum_{j=1,2} \left[\omega_j a_{j}^{\dag} a_{j} + \frac{\omega_{q_j}}{2} \sigma_{z}^{j} + i (-1)^{j} \, g_{0}(a_{j}^{\dag} \sigma_{-}^{j}- a_{j} \sigma_{+}^{j}) \right] + J(a_{1}^{\dag}a_{2} + a_{1} a_{2}^{\dag}) + \Omega \left[\sigma_{+}^{1} e^{-i\omega_{d}t} + \sigma_{-}^{1} e^{i\omega_{d}t} \right],
\end{eqnarray}
in which $a_j$ ($a^\dagger_j$) stands for the annihilation (creation) operator of the cavity modes, while $\sigma_{-}$ ($\sigma_{+}$) stands for the lowering (rising) operator of the qubits ($\vert \text{g} \rangle$ is the ground state while $\vert \text{e} \rangle$ is the excited state). Here, $\omega_j$ is the frequency of the cavity modes, $\omega_{q_j}$ is the frequency of the qubits, $g_0$ is the cavity-qubit coupling, $J$ is the hopping strength between the cavity modes, and $\Omega$ and $\omega_d$ are the amplitude and the frequency of the driving field, respectively (see Fig.~\ref{Fig1}). For this work, it is convenient to express the Hamiltonian in Eq. (\ref{Eqn2}) in the rotating frame with respect to the laser field. In such a case, the Hamiltonian is given by
\begin{figure*}[t!]
\centering
\includegraphics[width=1\linewidth]{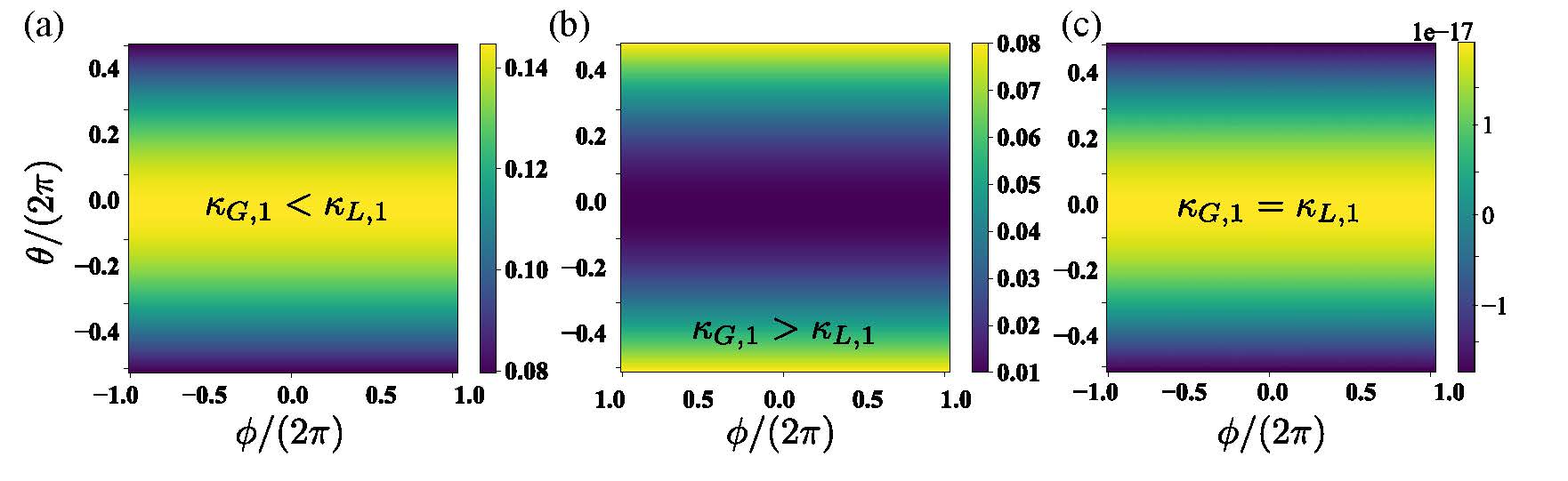}{\vspace{0.5em}}
 \caption{The limit cycle for the master system. Husimi Kano Q-representation function $Q(\theta,\phi)$ for the steady-state describing the two-level system for different dissipative configurations; (a) stands for the case where $\kappa_{G,1}/\kappa_{L,1}=10$, (b) $\kappa_{G,1}/\kappa_{L,1}=0.1$, and (c) $\kappa_{G,1}=\kappa_{L,1}$. The system parameters are given by $\Delta_{p,1}=10\kappa$, $\delta_{q,1}=0$, $g=0.5\kappa$, $J=10\kappa$.}
 \label{Fig2}
\end{figure*}
\begin{eqnarray} 
\label{Eqn2a}
\mathcal{H} =  \sum_{j=1,2} \left[\Delta_j a_{j}^{\dag} a_{j} + \frac{\delta_{q_j}}{2} \sigma_{z}^{j} + i (-1)^{j} \, g_{0}(a_{j}^{\dag} \sigma_{-}^{j}- a_{j} \sigma_{+}^{j}) \right] + J(a_{1}^{\dag}a_{2} + a_{1} a_{2}^{\dag}) + \Omega \sigma_{x}^{1},
\end{eqnarray}
where $\Delta_{j}=\omega_{j}-\omega_{d}$ and $\delta_{j}=\omega_{j}-\omega_{d}$ are the detuning of the $j$th mode and $j$th two-level system with respect to the driving frequency $\omega_{d}$, respectively. In what follows, we analyse the condition required to consider the two-level system composing the system as a self-sustained oscillator. 
\section{The limit cycle and phase locking}
In this section, we analyse the condition that the two-level system embedded in the master system must meet to describe a limit cycle. The limit cycle is defined as a stable trajectory which attracts nearby orbits. This cycle appears as a result of the competition between energy loss and energy gain on the system in the absence of an external signal \cite{arXiv.1812.10082}. Our figure of merit characterising the limit cycle of our master system correspond to the Husimi Kano Q-representation function \cite{book1} $Q(\theta,\phi)=\bra{\theta,\phi}{\rm{Tr}}_{f}[\rho_{{\rm st}}]\ket{\theta,\phi}/ 2\pi$, where $\ket{\theta,\phi}=(\cos\theta/2,e^{i\phi}\sin\theta/2)^{T}$ with $\theta \in\{0,\pi\}$ and $\phi\in\{0,2\pi\}$ is the $SU(2)$ generalized coherent state \cite{J.Phys.Math.33.3493}, and $\rho_{{\rm st}}$ corresponds to the steady state density matrix given by the following master equation
\begin{eqnarray}
\label{Eqn3}
\dot{\rho}(t) = -i[\mathcal{H}_{1},\rho] + \kappa_{L,1}\mathcal{D}[a_{1}]\rho + \kappa_{G,1}\mathcal{D}[a_{1}^{\dag}]\rho,
\end{eqnarray}
where $\mathcal{H}_{1}$ is the master system Hamiltonian expressed in the rotating frame with respect to the laser driving
\begin{eqnarray}
\label{Eqn4}
\mathcal{H}_{1} = \Delta_{1}a_{1}^{\dag}a_{1} + \frac{\delta_{q_1}}{2}\sigma^{1}_{z} + g_{0}(a_{1}^{\dag}\sigma_{-}^{1}+a_{1}\sigma_{+}^{1}).
\end{eqnarray} 

This representation permits to define an analogous to the phase space in spin systems. Furthermore, these states have the property that they precess over time according to $\ket{\theta,\phi}\rightarrow\ket{\theta,\delta_{q_1}t+\phi}$ \cite{arXiv.1904.11763}. This permits to define $\phi$ as the phase variable, which is essential in the synchronization phenomena. For any mixed two level system, the Husimi Kano Q-representation function takes the following form
\begin{eqnarray}
\label{Eqn5}
\bra{\theta,\phi}\rho\ket{\theta,\phi} =1+\cos\theta(2\rho_{00}-1) + 2 \sin\theta \mathfrak{R}[e^{-i\phi}\rho_{10}].
\end{eqnarray} 
Here, $\rho_{00}$ and $\rho_{01}$ correspond to the matrix element of the steady-state density matrix, which provides information about the population and the coherences, respectively. Furthermore, $\mathfrak{R}[z]$ stands for the real part of the complex number $z$. From Eq. (\ref{Eqn5}) it is easy to see that for steady-state of the form $\rho=p_{0}\ketbra{0}{0} + (1-p_{0})\ketbra{1}{1} $ (absence of coherence), the $Q$ representation function is phase free i.e., the representation only depends on $\theta$ as shown in Fig. \ref{Fig2}. In such a case, the dissipative source of energy produced by the master equation on the system does not prefer any particular phase on the steady-state. As a result, the $Q(\theta,\phi)$ representations have diffusive shape for both cases, i,e.,  $\kappa_{G,1}>\kappa_{L,1}$ (Fig. \ref{Fig2}(a)), and $\kappa_{G,1}<\kappa_{L,1}$ (Fig. \ref{Fig2}(b)), respectively. Moreover, the maximal/minimal value of $Q$ is achieved at $\theta=0$, which corresponds to a state precesing on the $z$ axis. On the other hand, Fig.~\ref{Fig2}(c) shows that equal loss/gain energy ratios in the $Q$ representation are zero for any value of $\{\theta,\phi\}$. This occurs because the steady state corresponds to a maximally mixed state which does not precess in the $z$ axis. These facts constitute the required condition to achieve the limit cycle on the two-level system composing the master system.\\
\begin{figure*}[t!]
\centering
\includegraphics[width=1\linewidth]{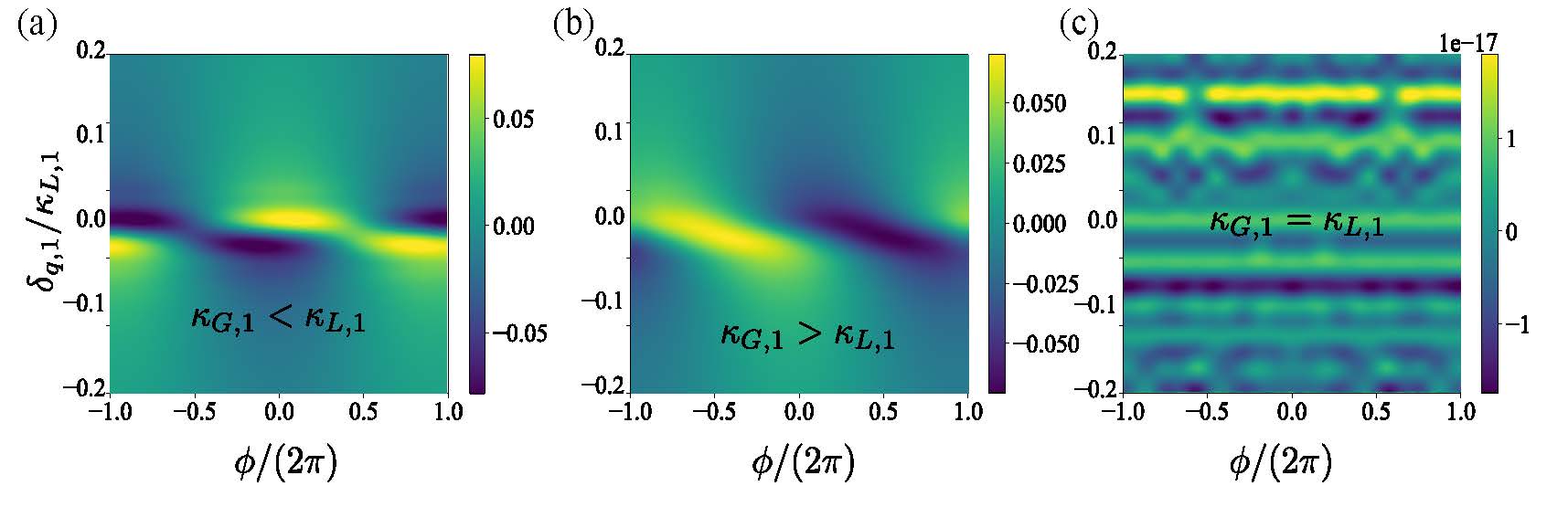}{\vspace{0.5em}}
 \caption{Phase distribution $\mathcal{P}(\phi,\rho) $ as a function of the two-level system detuning $\delta_{A,1}$; (a) stands for the case where $\kappa_{G,1}/\kappa_{L,1}=10$, (b) $\kappa_{G,1}/\kappa_{L,1}=0.1$, and (c) $\kappa_{G,1}=\kappa_{L,1}$. The system parameters are given by $\Delta_{p,1}=10\kappa$, $\Omega=0.01\kappa_{L,1}$, $g=0.5\kappa$, $J=10\kappa$.}
 \label{Fig3}
\end{figure*}
Another feature to be considered in the description of the self-sustained oscillator corresponds to the phase-locking observed when the system interacts with an external signal. Phase locking can be analyzed in terms of a phase distribution containing all the information about both the phase and the steady state of the system. This phase distribution can be written as the polar average of the Husimi $Q$ representation \cite{Phys.Rev.A.99.043804} as follows
\begin{eqnarray}
\label{Eqn6}
\mathcal{P}(\phi,\rho) = \int_{0}^{\pi}\tilde{Q}(\theta,\phi)\sin\theta d \theta-\frac{1}{2\pi}.
\end{eqnarray}
Here, $\tilde{Q}(\theta,\phi)$ corresponds to the Husimi Kano Q-representation function of the steady state given by the master equation represented in the rotating frame with respect to the classical laser
\begin{eqnarray}
\label{Eqn7}
\dot{\rho}(t) = -i[\mathcal{H}_{1},\rho] -i[\Omega \sigma^{1}_{x},\rho]+ \kappa_{L,1}\mathcal{D}[a_{1}]\rho + \kappa_{G,1}\mathcal{D}[a_{1}^{\dag}]\rho,
\end{eqnarray}
which is nothing more than the master equation on Eq. (\ref{Eqn1}) with a driving term. On the other hand, $\mathcal{P}(\phi,\rho) $ is zero when the oscillator is phase free. This condition resembles the similar behaviour observed in a noisy classical limit-cycle oscillator, where the oscillator has a uniform phase distribution \cite{Phys.Rev.A.99.043804}. 

\begin{figure*}[t!]
\centering
\includegraphics[width=1\linewidth]{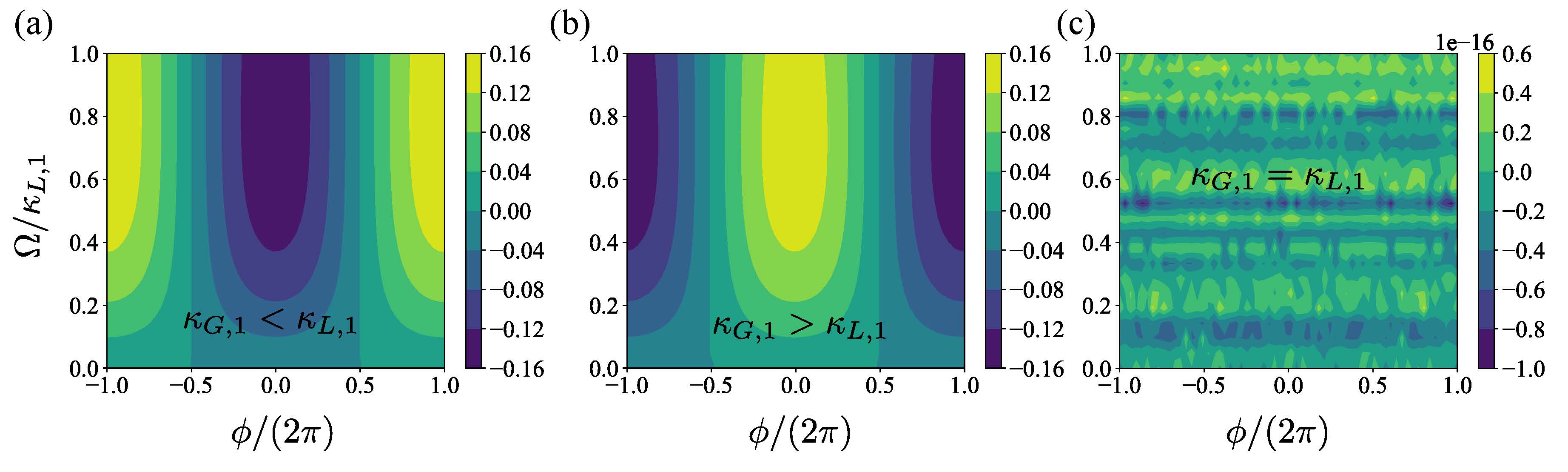}{\vspace{0.5em}}
 \caption{Phase distribution $\mathcal{P}(\phi,\rho) $ as a function of the external signal strength $\Omega$; (a) stands for the case where $\kappa_{G,1}/\kappa_{L,1}=10$, (b) $\kappa_{G,1}/\kappa_{L,1}=0.1$, and (c) $\kappa_{G,1}=\kappa_{L,1}$. The system parameters are given by $\Delta_{p,1}=10\kappa$, $\delta_{q,1}=0$, $g=0.5\kappa$, $J=10\kappa$.}
 \label{Fig4}
\end{figure*}

Figure \ref{Fig3} shows the phase distribution as a function of the detuning $\delta_{q,1}$ for several dissipative configurations. As expected, when one of the dissipative rates dominates the phase-locking on the two-level system emerges (see Fig. \ref{Fig3}(a) and Fig. \ref{Fig3}(b)). In these cases, the phase distribution reaches its optimal value in a vicinity of the zero detuning case i.e., $\delta_{q,1}\in\{-0.1,0.1\}~\kappa_{L,1}$. These optimal values are related to the phase, or anti-phase locking observed between the signal and the two-level system. Finally, for equal loss/gain rate, there is no phase locking. In fact, as $\mathcal{P}(\phi,\rho) $ is zero in all the parameter space we say that synchronisation has not achieved and the qubit is free-phase. On the other hand, Fig. \ref{Fig4} shows $\mathcal{P}(\phi,\rho)$ as a function of the external signal strength $\Omega$. In this case, we achieve phase-locking on the two-level system when one of the dissipative scales dominates. Moreover, the optimal value of the phase distribution gives us information concerning with the phase (Fig. \ref{Fig4}(a)) or anti-phase (Fig. \ref{Fig4}(b)) nature of the synchronisation, respectively. For the case where loss and gain ratios are equal, we observe again that there is no syncrhonization and the two-level system is free-phase.\\
Finally, due to we can define a limit cycle on the master system, as well as show phase-locking between the signal and the two-level system, it is possible to say that the two-level system composing the master system (also for the slave system) constitutes a \textit{self-sustained} oscillator. 
\section{The effective model}
In our work, it is possible to distinguish two subsystems; \textit{the master system} and \textit{the slave system} \cite{Pikovsky2003} which are constituted by the first, and second qubit-cavity system, respectively. The former characterizes for having an independent movement provided by the signal applied to it. On the contrary, the slave system has a non-free movement as a result of the cavity-cavity interaction. To understand how the cavity-cavity interaction allows us to achieve synchronization between the two level systems on the master and slave systems, respectively, it is convenient to write the cavity interaction in the normal modes basis i.e, $b_{1} = (a_{1}-a_{2}) / \sqrt{2}$ and $b_{2} = (a_{1}+a_{2}) / \sqrt{2}$. Notice that this transformation is valid when the cavity modes are equal i.e., $\omega_{1}=\omega_{2}=\omega$, obtaining the following master equation
\begin{eqnarray}
 \label{Eqn8}
\dot{\rho}=-i[\mathcal{H}_{{\rm nm}},\rho ] + \sum_{j=1,2} \big[\kappa_{L,j}\mathcal{D}[b_{j}]\rho + \kappa_{G,j}\mathcal{D}[b_{j}^{\dag}]\rho\big],
\end{eqnarray}
where $\mathcal{H}_{{\rm nm}}$ is the Hamiltonian in Eq. (\ref{Eqn1}) expressed in term of the normal mode
\begin{eqnarray}
 \label{Eqn9}
\mathcal{H}_{{\rm nm}} = (\omega - J) b_{1}^{\dag} b_{1} + (\omega + J) b_{2}^{\dag} b_{2} +  \sum_{j=1,2} \frac{\omega_{q_j}}{2} \sigma_{z}^{j}  - ig_{0}\bigg[b_{1}^{\dag}(\sigma_{-}^{1} + \sigma_{-}^{2})  + b_{2}^{\dag}(\sigma_{-}^{1} - \sigma_{-}^{2}) + \rm{H.c}\bigg] + \Omega[\sigma_{+}^{1}e^{-i\omega_{d}t} + \sigma_{-}^{1}e^{i\omega_{d}t} ] ,
\end{eqnarray}
Considering the case of equal qubit frequencies, we set the frequency of the external driving field quasi-resonant to the atomic transition and to the high-frequency normal mode ($\omega_{d} \approx \omega_{q} \approx \omega + J$). In the interaction picture with respect to $\mathcal{H}_{0} = (\omega - J) b_{1}^{\dag} b_{1} + (\omega + J) b_{2}^{\dag} b_{2} +  \sum_{j=1,2} \frac{\omega_{q}}{2} \sigma_{z}^{j}$, we have
\begin{eqnarray} 
\label{Eqn10}
\bar{\mathcal{H}}_{{\rm nm}} =  \Omega\sigma_{x}^{1} - ig_{0}\bigg[b_{1}^{\dag}S_{-}e^{2iJt} + b_{2}^{\dag}Q_{-} + \rm{H.c}\bigg],
\end{eqnarray}
where $S_{\pm}=(\sigma_{\pm}^{1}+\sigma_{\pm}^{2})/\sqrt{2}$, and $Q_{\pm}=(\sigma_{\pm}^{1}-\sigma_{\pm}^{2})/\sqrt{2}$ are collective spin operators. By considering $2\vert J \vert \gg \vert g_{0} \vert$, we see that the second term on Eq. (\ref{Eqn10}) is highly oscillating. In order to eliminate this term, we apply the following unitary transformation over the master equation in Eq. (\ref{Eqn1}) 
\begin{eqnarray}
\label{Eqn11}
\mathcal{U} = e^{-i\omega_{d}\sum_{j}(b_{j}^{\dag} b_{j}+ \sigma_{z}^{j}/2)t} \times e^{\frac{g_{0}}{2J}(S_{-}b_{1}^{\dag} - S_{+}b_{1})} \times e^{i\omega_{d}\sum_{j}(b_{j}^{\dag} b_{j}+ \sigma_{z}^{j}/2)t}.
\end{eqnarray}
Neglecting highly-oscillating terms, the dynamics up to second order in $g_{0}/2J$ is given by
\begin{eqnarray} 
\label{Eqn12}
\dot{\rho}=-i[\tilde{\mathcal{H}}_{{\rm nm}},\rho ] + \sum_{j=1,2} \big[\kappa_{L,j}\mathcal{D}[b_{j}]\rho + \kappa_{G,j}\mathcal{D}[b_{j}^{\dag}]\rho\big] +  \left(\frac{g_{0}^{2}}{2J} \right)\sum_{j=1,2} \big[\kappa_{L,j}\mathcal{D}[S_{-}]\rho + \kappa_{G,j}\mathcal{D}[S_{+}]\rho\big],
\end{eqnarray}
where $\tilde{\mathcal{H}}_{{\rm nm}}$ is the effective normal mode Hamiltonian in the Schr\"odinger picture
\begin{eqnarray}\nonumber
\label{Eqn13}
\tilde{\mathcal{H}}_{{\rm nm}} &=& \left(\omega - J + \frac{g_{0}^{2}}{2J}\frac{S_{z}}{2} \right) b_{1}^{\dag} b_{1} + (\omega + J) b_{2}^{\dag} b_{2} +  \left( \omega_{q} + \frac{g_{0}^{2}}{4J} \right) \frac{S_{z}}{2} \\
&-& ig_{0}(b_{2}^{\dag} Q_{-}-b_{2}Q_{+}) - \frac{g_{0}^{2}}{4J}(\sigma_{+}^{1} \sigma_{-}^{2} + \sigma_{-}^{1}\sigma_{+}^{2}) + \Omega \left[\sigma_{+}^{1} e^{-i\omega_{d}t } + \sigma_{-}^{1} e^{i\omega_{d}t } \right].
\end{eqnarray}
Then, we observe that the nonresonant term induces shifts in the low-frequency normal mode and qubit bare energies, and an effective dissipative term for the qubits (collective dissipation). Assuming $2\vert J\vert \gg \kappa_{L,j} \gg \{\vert g_{0}/2\vert$, $\vert \Omega g_{0}/4J \vert\}\gg \kappa_{G,j}$ and adjusting $\omega_{q} = \omega + J - g_{0}^{2}/4J$ (effective resonance between the qubits and the high-frequency normal mode), the field variables can be adiabatically eliminate \cite{zoller}, so that the reduced dynamics for the qubits, in the interaction picture, is given by
\begin{eqnarray} \label{Eqn15}
\dot{\rho}&=&-i[\tilde{\mathcal{H}}_{{\rm nm}},\rho ] +\frac{g_{0}^{2}}{\kappa_{L}}\mathcal{D}[Q_{-}] + \frac{g_{0}^{2}}{\kappa_{G}}\mathcal{D}[Q_{+}] +  \left(\frac{g_{0}^{2}}{2J} \right)\bigg[\kappa_{L}\mathcal{D}[S_{-}]\rho + \kappa_{G}\mathcal{D}[S_{+}]\rho\bigg]\\
\mathcal{H} &=& - \frac{g_{0}^{2}}{4J}(\sigma_{+}^{1} \sigma_{-}^{2} + \sigma_{-}^{1}\sigma_{+}^{2}) + \Omega \sigma_{x}^{1}.
\end{eqnarray}
The approximation allows us to identify that the high-frequency normal mode effectively acts as a common reservoir that couples the qubits. Such effective coupling will be responsible for the onset of synchronization.\\
\section{Numerical Result}

\begin{figure*}[t!]
\centering
\includegraphics[width=1\linewidth]{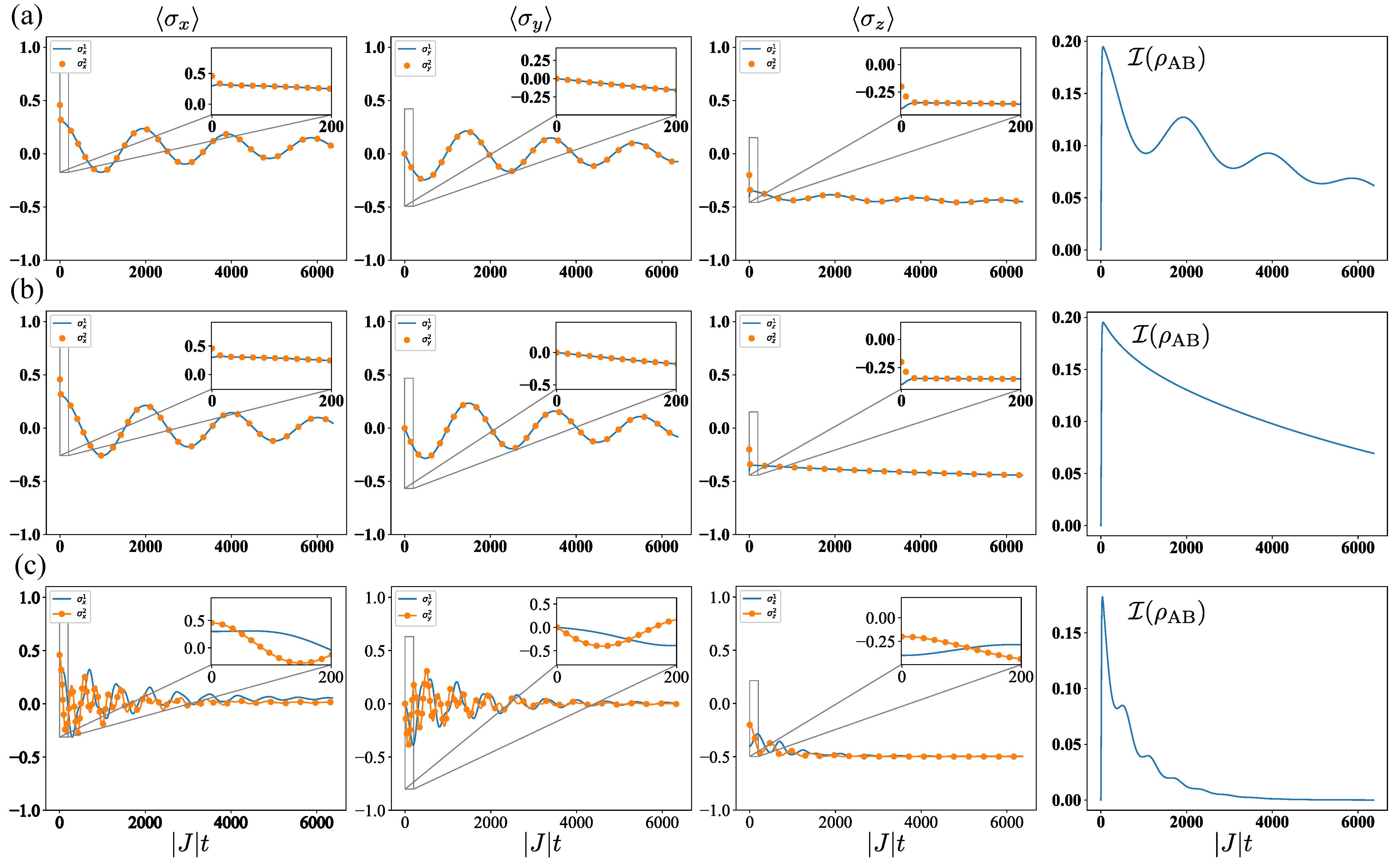}{\vspace{0.5em}}
 \caption{Time evolution of the expectation values of the Pauli operators corresponding to qubit $1$ (light color) and qubit $2$ (dark color), and the mutual information between the qubits. Here, we consider $\kappa_{L,1}=\kappa_{L,2}=\kappa$, $\kappa_{G,1}=\kappa_{G,2}=0.01\kappa$, $J = -10\kappa$, $g_{0}=\kappa/2$, and $\Delta_1 = \omega_1-\omega_d = -J$, together with (a) $\Delta_2 = \omega_2-\omega_d = \Delta_1$, $\delta_1 =\delta_2 = 0$ and $\Omega = 5\times10^{-4}\kappa$, (b) the same of (a) except for $\Omega = 0$, (c) $\Delta_2 = \Delta1$, $\delta_1 =0\kappa, \delta_2 = -0.03 \kappa$, and $\Omega = 1\times10^{-3}\kappa$. Initially, the cavities are in the vacuum state while the qubits are in state $\vert \psi(0) \rangle = (\sqrt{0.9} \vert \text{g} \rangle + \sqrt{0.1} \vert \text{e} \rangle) \otimes (\sqrt{0.7} \vert \text{g} \rangle + \sqrt{0.3} \vert \text{e} \rangle)$. We observe a total quantum synchronization in (a), i.e., all Pauli operators are synchronized ($\Xi=2$ $-$ maximum degree of quantumness). In (b), we have a partial quantum synchronization ($\Xi=1$), since just $\sigma_x$ and $\sigma_y$ are synchronized, while we have a classical synchronization ($\Xi=0$) in (c), because only $\sigma_z$ is synchronized. In every case, we observe synchronization even in the absence of correlations between the qubits in the steady state.}
 \label{Fig5}
\end{figure*}
A numerical simulation of the evolution of the expectation values of Pauli operators according to Eq.~\eqref{Eqn1}, in the rotating frame with laser frequency, for three sets of parameters is depicted in Fig.~\ref{Fig5}, as well as the quantum mutual information between the qubits. The quantum mutual information $I$ quantifies how much the knowledge about the system A gives information about the system B. It is defined as $I(\rho_{AB}) = S(\rho_A) + S(\rho_B) - S(\rho_{AB})$, where $S(\rho) = -\text{Tr}(\rho\log\rho)$ is the von Neumann entropy, with $\rho_A = \text{Tr}_B(\rho_{AB})$ and $\rho_B = \text{Tr}_A(\rho_{AB})$. In Fig.~\ref{Fig5}(a), we observe that the external driving field induces a total quantum synchronization of the qubits (maximum degree of quantumness $\Xi=2$), since all Pauli operators get almost immediately synchronized, even in amplitude. If the driving field is off, see Fig.~\ref{Fig5}(b), the qubits spontaneously synchronize. However, it is a partial quantum synchronization since just $\sigma_x$ and $\sigma_y$ get synchronized ($\Xi=1$), while $\sigma_z$ for both qubits exhibits a pure exponential decay. In Fig.~\ref{Fig5}(c), we consider a case in which the external driving field can only induce (anti-)synchronization in $\sigma_z$, i.e., a classical synchronization ($\Xi=0$). Although it is not depicted in Fig.~\ref{Fig5} for the sake of clarity, the approximate reduced dynamics [Eq.~\eqref{Eqn15}] reproduces the exact numerical results with excellent agreement, ratifying the statement made in the previous paragraph.

In Ref.~\cite{Ameri2015}, the authors propose the use of quantum mutual information as a steady-state order parameter for signaling the presence or the absence of quantum synchronization, claiming that it is well defined for every bipartite quantum state and does not depend on the particular details of the system. They suggest that synchronized systems should converge to a steady state having large mutual information. The intuition behind this proposal is that, under a quantum dynamics, quantum correlations tend to emerge. Surprisingly, every case shown in Fig.~\ref{Fig5} exhibits synchronization even though the amount of mutual information embedded on the two-level systems bipartition is small. Therefore, as already noticed for entanglement  \cite{Zhirov2009,Lee2014,Manzano2013,Mari2013,Ameri2015}, there is not a one-to-one correspondence between correlations in the steady state and synchronization, and this relation strongly depends on the specific details of the system. From our point of view, quantum mutual information is a signal of quantum synchronization, but the opposite is not true, and a quantum dynamics can yield quantum synchronization without generating a high mutual information. \\

\begin{figure*}[!t]
\centering
\includegraphics[width=1\linewidth]{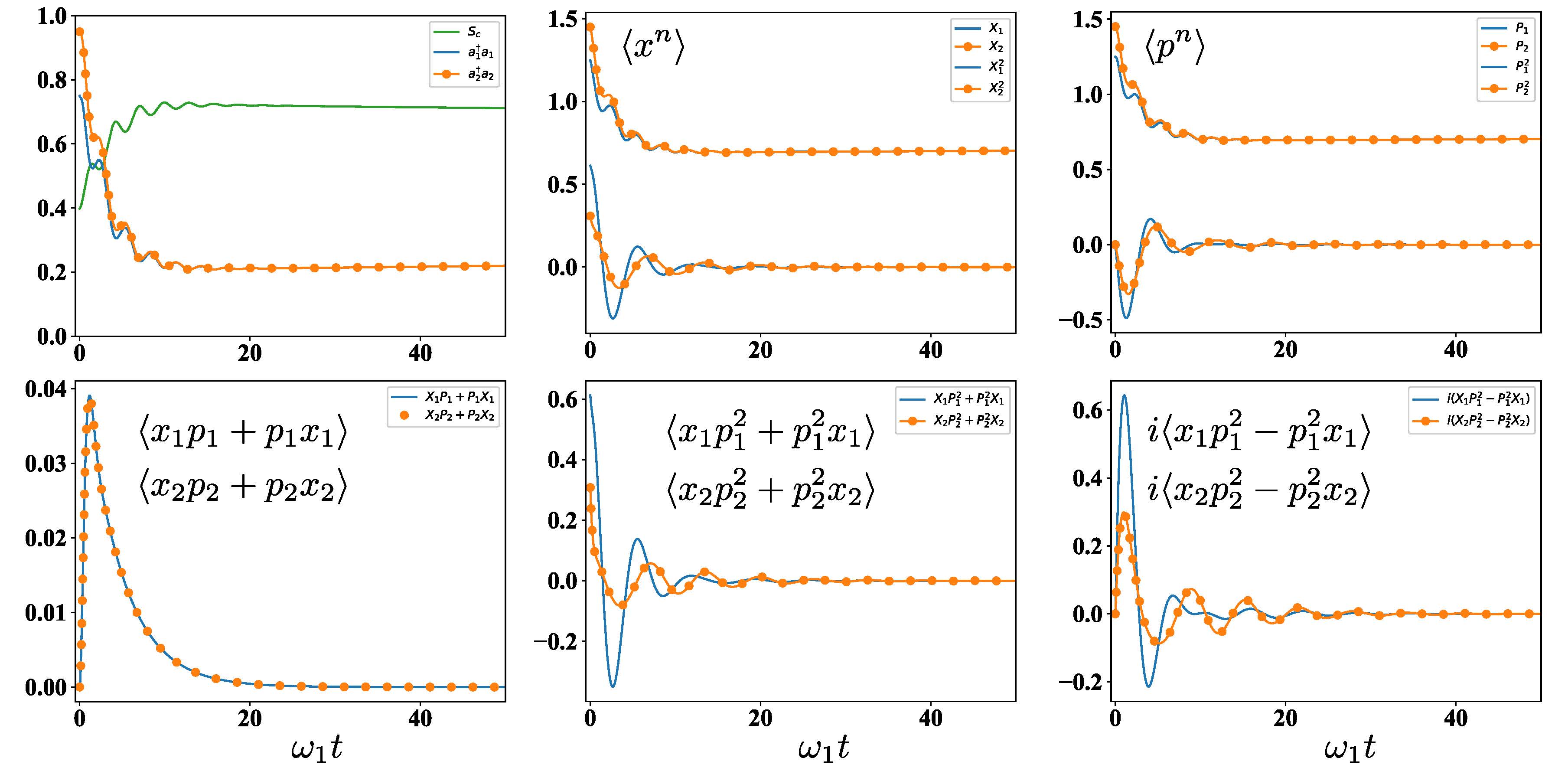}{\vspace{0.5em}}
\caption{Time evolution of $S_c$, the average number of photons, and the average of the lowest moments of the oscillators. Here, we consider $\kappa_{L,1} = \kappa_{L,2} = 2\omega_2 = 2\omega_1$, $J = 0.5\omega_1$ and $\kappa_{G,1} = \kappa_{G,2} = 0.001 \omega_1$. The initial state is $\vert \psi(0) \rangle = (\sqrt{0.25} \vert \text{0} \rangle + \sqrt{0.75} \vert \text{1} \rangle) \otimes (\sqrt{0.05} \vert \text{0} \rangle + \sqrt{0.95} \vert \text{1} \rangle)$, where $\vert n \rangle$ are Fock states.}\label{vdp}
\end{figure*}

Distinct resonator frequencies can be created by means of different resonator lengths, and the interaction between different cavity modes, known as mode mixing, occurs via tunneling of photons. The corresponding hopping term which connects both resonators in the Hamiltonian, can be implemented using a superconducting quantum interference device (SQUID) made of a superconducting loop interrupted by two Josephson junctions (JJs), provided that $\vert J \vert / \omega \ll 1$ \cite{Felicetti2014}. Each of the resonators, on the other hand, contains a superconducting qubit, which can be coupled to cavities well beyond the value $g_0/ \omega \approx 0.01$ that we require, as it has been reported very recently by Bosman \textit{et al.}~\cite{Bosman2016}. Additionally, they can reach coherence times as high as $100 \mu$s~\cite{Koch2007, Paik2011, Rigetti2012}, while the plotted amount of time in Fig.~\ref{Fig5}, on the other hand, corresponds to the order of $1 \mu$s, which means that the observation of hundreds of oscillations is available.  Finally the driving field on one of the qubits is implemented via a coherent microwave source~\cite{Mallet2009}.

\section{Coupled Van der Pol Oscillators}
We finally focus on a model made of two coupled quantum van der Pol Oscillators oscillators. We introduce an in-phase synchronizing Hamiltonian
\begin{eqnarray}
\label{Eqn16}
\mathcal{H}_{\text{vdP}} &= \omega_1 a_1^\dagger a_1 + \omega_2 a_2^\dagger a_2 + iJ(a_1^\dagger a_2^\dagger - a_1 a_2).
\end{eqnarray}
The dynamics of the coupled vdP oscillators is given by the following master equation is given by~\cite{Lee2013}

\begin{eqnarray}
\label{Eqn17}
\dot{\rho} = -i[H_{\text{vdP}},\rho ] + \sum_{j=1,2} \kappa_{G,j}\mathcal{D}[a_{j}^{\dag}]\rho + \kappa_{L,j}\mathcal{D}[a_{j}^{2}]\rho,
\end{eqnarray}

In this model, each oscillator gains one photon at rate $\kappa_{G,j}\langle a_{j}a_{j}^{\dag}\rangle$, and losses two photons at rate $\kappa_{G,j}\langle (a_{j}^{\dag})^{2}a_{j}^{2}\rangle$. These processes resemble the non-linear and quadratic damping observed in the classical vdP oscillator in absence of external signal \cite{Lee2013}. The lowest moments for this model are depicted in Fig.~\ref{vdp}, showing a transient oscillatory behavior in which synchronization is observed in all of them, except for $\langle xp + px \rangle/2$. The quantum synchronization figure of merit $S_c(t)  := \langle x_{-}(t)^2 + p_{-}(t)^2 \rangle^{-1} \le1$, introduced by A. Mari \textit{et al}.~\cite{Mari2013}, is also plotted, where $x_{-}(t) := [x_{1}(t)-x_{2}(t)]/\sqrt{2}$ and $p_{-}(t) := [p_{1}(t)-p_{2}(t)]/\sqrt{2}$, with $x_{j} =( a_j + a_{j}^{\dagger})/\sqrt{2}$ and $p_{j} =-i( a_j - a_{j}^{\dagger})/\sqrt{2}$ the canonical variables of each oscillator. These simulations suggest that almost full degree of quantum synchronization is attained. However, the existence of infinite moments gives rise to the theoretical challenge of proving that higher moments are actually synchronized, since the numerical approach is limited. This, therefore, means that there are still open questions in the case of continuous variables which should be addressed in future research. 

\section{Conclusions}
Summarizing, in this work we have proposed a quantifier of the quantumness of a quantum synchronization process based on quantum information techniques. Indeed, we define the degree of quantumness in terms of the number of synchronized non-commuting observables. This approach is different to the previous works, since we are not proposing another measure of synchronization, but of the quantumness of the generated quantum synchronization. We study in detail the case of finite-dimensional systems, illustrating it with two cavity-qubit systems, in which we show that all possible degrees of quantumness may be reached for the qubits. Additionally, we show that this setup is feasible in superconducting circuits with current technology. Finally, we analyze the case for continuous variables, illustrating it with two quantum van der Pol oscillators, where we show that there are still open questions which should be addressed in the future.

{\setlength{\parindent}{0pt}

\section*{AUTHOR CONTRIBUTIONS}
H.E., as the first author, has been responsible together with M.S. for the development of this work. M.S., supported by D.Z.R., and F.A.C.L., have made the mathematical demonstrations, carried out calculations, and provided examples. H.E. and D.Z.R. have performed the numerical simulations. Finally, M.S. suggested the seminal ideas and, together with E.S., supervised the project throughout all stages. All authors have carefully proofread the manuscript.

\section*{ACKNOWLEDGMENT}
The authors thank L. Garc\'ia-\'Alvarez, I. L. Egusquiza, R. Zambrini and F.~Deppe for fruitful discussions. D.Z.R. acknowledges support from S\~{a}o Paulo Research Foundation (FAPESP) Grants No.~2013/23512-7 and No.~2014/24576-1. While M.S. and E.S. are grateful for the funding from Spanish MCIU/AEI/FEDER (PGC2018-095113-B-I00), Basque Government Grant No.~IT986-16, the projects QMiCS (820505) and OpenSuperQ (820363) of the EU Flagship on Quantum Technologies and the EU FET Open Grant Quromorphic (828826). This work is supported by the U.S. Department of Energy, Office of Science, Office of Advanced Scientific Computing Research (ASCR) quantum algorithm teams program, under field work proposal number ERKJ333.

\section*{ADDITIONAL INFORMATION} 
The authors declare no competing financial interests.

\end{document}